# Spatial Matching of 2D Mammography Images and Specimen Radiographs: Towards Improved Characterization of Suspicious Microcalcifications

Noor Nakhaei*[a,b], Chrysostomos Marasinou[a,b], Akinyinka Omigbodun[a,b], Nina Capiro[c], Bo Li[c], Anne Hoyt[c], William Hsu[a,b]

[a]Department of Computer Science, University of California, Los Angeles, CA USA, [b]Medical & Imaging Informatics, Department of Radiological Sciences, University of California, Los Angeles, CA USA, [c]Breast Imaging Section, Department of Radiological Sciences, University of California, Los Angeles, CA USA

## ABSTRACT

Accurate characterization of suspicious microcalcifications is critical to determine whether these calcifications are associated with invasive disease. Our overarching objective is to enable the joint characterization of microcalcifications and surrounding breast tissue using mammography images and digital histopathology images. Towards this goal, we investigate a template matching-based approach that utilizes microcalcifications as landmarks to match radiographs taken of biopsy core specimens to groups of calcifications that are visible on mammography. Our approach achieved a high negative predictive value (0.98) but modest precision (0.66) and recall (0.58) in identifying the mammographic region where microcalcifications were taken during a core needle biopsy.

## 1. INTRODUCTION

Breast cancer is the most common invasive cancer and the second leading cause of death in women [1]. Routine screening mammography has the potential to detect breast cancer in its earliest stage, prior to it becoming a potentially lethal invasive breast cancer. Ductal carcinoma in situ (DCIS) is breast cancer confined to the milk duct and is the earliest stage of breast cancer. DCIS may present with mammographically visible microcalcifications. However, making a diagnosis of DCIS can be challenging since most microcalcifications seen on a mammogram are benign or non-cancerous. So, for every woman who undergoes screening mammography followed by breast biopsy and receives a new diagnosis of breast cancer, approximately two additional women will undergo a benign breast biopsy. Although modern percutaneous core needle biopsy is safe, unnecessary biopsies are anxiety-provoking and are considered a risk associated with breast cancer screening.

Microcalcifications with suspicious morphology (e.g., amorphous) are particularly challenging, given that the difficulty of characterizing them and the fact that the decision threshold to biopsy them is low.

**Figure 1** depicts the sequence of events when indeterminate screen-detected mammographic calcifications are identified. The patient is recalled for further evaluation with diagnostic mammography and magnification of the calcifications. The magnified diagnostic views allow the radiologist to determine if the mammographic calcifications are suspicious, probably benign, or benign. Suspicious calcifications warrant biopsy. Probably benign calcifications should be followed with repeat short interval imaging in 6-month to assess for change or stability, those with benign calcifications return to routine mammographic screening. Women who are recalled from screening mammography for additional evaluation experience anxiety associated with the need to return for additional imaging and, if biopsy is indicated, anxiety about the procedure and results. Unnecessary benign biopsies are known as false positives. It is desirable to minimize these unnecessary biopsies while maintaining a high mammographic sensitivity (i.e. not missing any cancers).

Towards addressing the challenge of managing suspicious microcalcifications, our overall goal is to develop a computer-aided diagnosis algorithm that quantitatively analyzes the morphology and distribution of microcalcifications on diagnostic mammograms to distinguish between cases that should undergo biopsy versus short term follow-up imaging. Given that microcalcifications are a byproduct of biological processes that may indicate the presence of invasive cancer, we wish to relate the mammographic appearance of microcalcifications to the tissue and cellular structure that is observable under the microscope. However, precisely pinpointing the region in the mammogram where a biopsy specimen was taken is challenging. One potential approach is to utilize the specimen radiographs of biopsy cores that are taken clinically to ensure that the targeted region of microcalcifications were collected during biopsy. In this work, we investigate the feasibility of using the shape and distribution of microcalcifications in specimen radiographs to identify a region in the mammography image that appears to show similar microcalcifications. We posit that the ability to link a specific region in the mammography image to the tissue imaged in the specimen radiograph will enable more precise radiology-pathology correlation.

## 2. METHODS

### 2.1 Dataset

Data were collected retrospectively following an institution review board approved protocol from patients seen at a single academic medical center. The dataset consisted of diagnostic mammograms from 80 patients. Each patient had two mammogram views: magnified craniocaudal (CC) view and magnified 90-degree mediolateral or lateromedial (ML/LM) view. The magnified CC-view images were used for training, as suggested by clinicians, because microcalcifications are more visible on CC versus other views. All other views were used for testing, and 51 CC and ML/LM views were for testing. All images were acquired using Hologic Selenia full-field digital mammography equipment at 0.070 mm per pixel resolution and 12-bit grayscale. For each case, specimen radiographs of the biopsied tissue were also obtained by placing the tissue cores into a plastic tray and imaging the specimen via x-ray. A breast fellowship-trained, board-certified radiologist (BL) and a breast fellow (NC) reviewed all of the full views, magnified views, and the specimen x-rays and annotated individual microcalcifications. The open-source medical image viewer Horos was utilized to create the annotations, marking the spatial location of visible microcalcifications that were the biopsy target with a single point. The smallest bounding box containing the points annotated on the mammograms was used to denote the reference biopsied region to which the results of the template matching approach was compared.

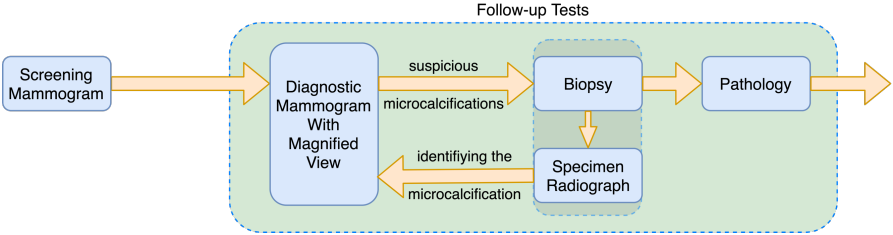

**Figure 1.** The general process for diagnosing suspicious microcalcifications. The magnification view obtained during the diagnostic mammogram and specimen radiographs acquired post-biopsy are the focus of this work.

## 2.2 Overall approach

The process for finding the region on the mammogram associated with the calcifications in the specimens is illustrated in **Figure 2**. Briefly, magnification views of the diagnostic mammogram and specimen radiograph are automatically segmented using the same approach. The diagnostic mammogram is then divided into non-overlapping patches. A clustering algorithm is applied to the segmented specimen radiographs to group related calcifications. These groups are used as templates and matched against each patch within the mammogram. The output of the template matching process is a score that is used to determine the likely patch from which the microcalcification group in the specimen radiograph came. The following sections describe each step in detail.

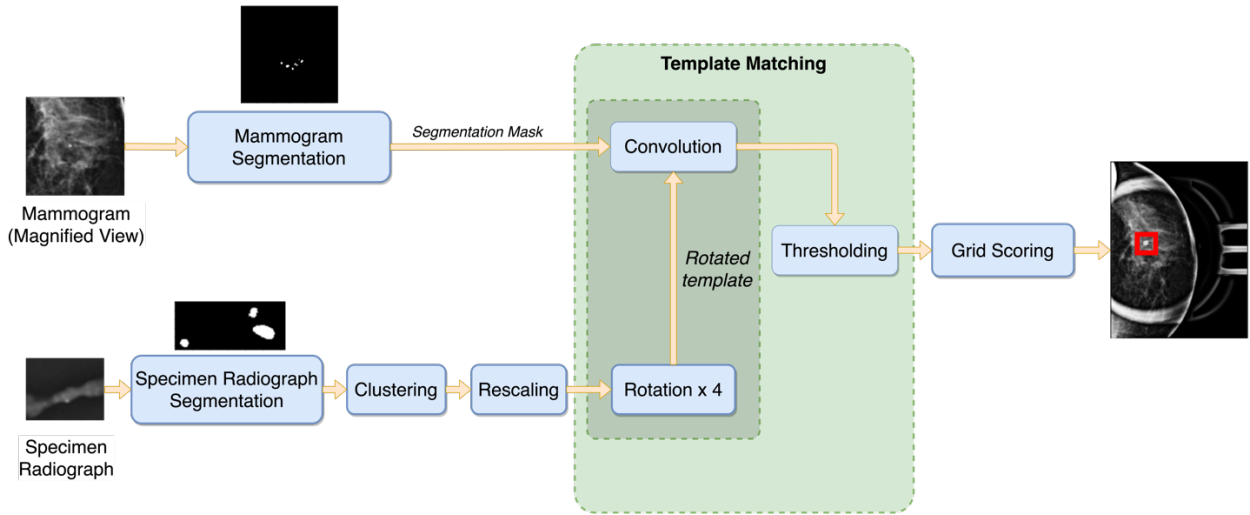

**Figure 2.** Overall approach for spatially matching microcalcifications imaged in the diagnostic mammogram and specimen radiograph.

## 2.3 Microcalcification segmentation

Calcifications captured in magnification views were detected and segmented by an algorithm developed by our team [2]. A sample result is shown in **Figure 3a**. The method consists of two stages: (1) bright candidate objects are delineated using difference-of-Gaussians with Hessian analysis and (2) a convolutional regression model is applied to choose the candidate objects corresponding to calcifications. The calcifications on the specimen radiograph are segmented using the same method; an example result is shown in **Figure 3b.** Two image masks are generated: one of the microcalcifications from the magnified view, and the other of the microcalcifications from the specimen radiograph.

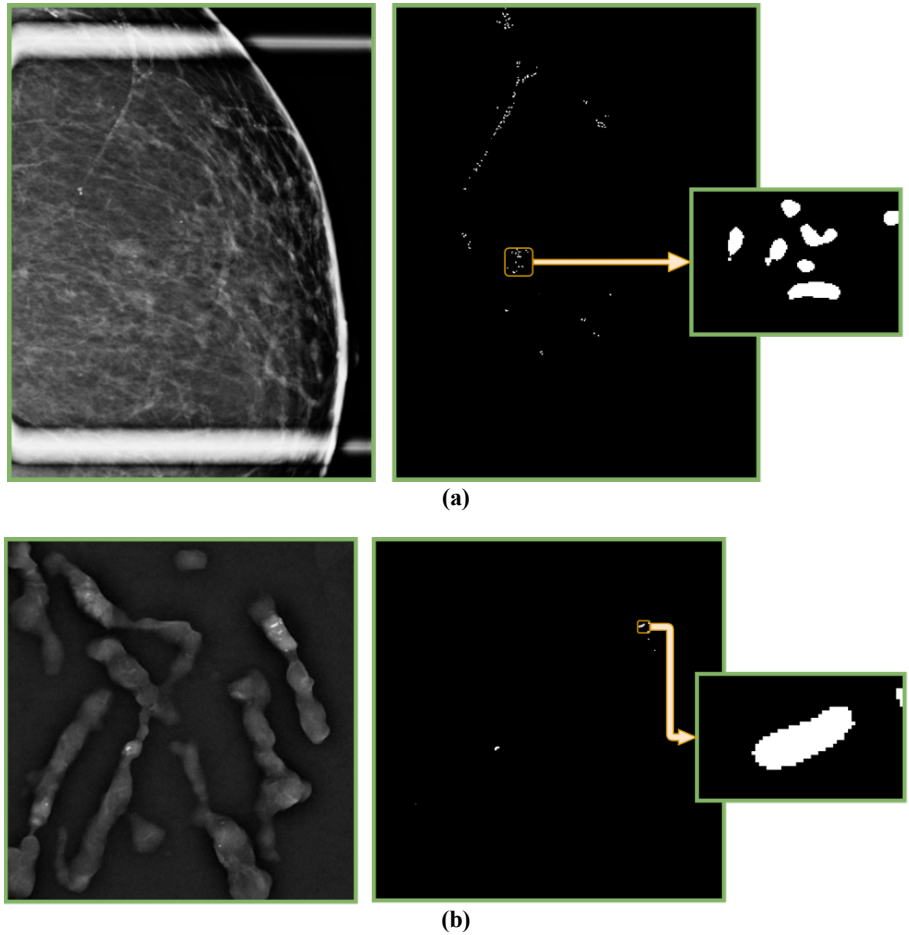

**Figure 3.** Examples of segmented microcalcifications **(a)** in the magnification view of the diagnostic mammogram and **(b)** the specimen radiograph.

## 2.4 Clustering microcalcifications

To provide a more robust way to match specimen radiographs to mammograms, we investigated ways to identify groups of microcalcifications observed on the specimen radiographs that could be used as landmarks. To generate these groups, segmented microcalcifications on the specimen radiograph were clustered using an unsupervised clustering approach called Density-Based Spatial Clustering of Application with Noise (DBSCAN) [3]. The objects were mapped to their centroid locations (points). Utilizing DBSCAN, points were grouped based on their neighborhood density and a minimum number of points within the group. Points that did not meet the minimum threshold for a group were labeled as outliers. A bounding box that encompasses the largest identified cluster of calcifications was cropped from the image and used as the template. Given that specimen radiographs may be magnified during acquisition, a field in the DICOM header called Estimated Radiographic Magnification Factor, was used as a scaling factor. Relative orientations of the calcifications may vary between the mammogram and the specimen radiograph. As such, from each template, three additional templates were generated by rotating the original template by 90, 180, and 270 degrees. All four templates were compared to the mammographic calcification mask.

The mammogram was divided into a grid of 300x300 non-overlapping patches. Patches that overlapped with the bounding box annotations by the human readers where biopsies were taken were considered positive patches. All other patches were considered negative patches.

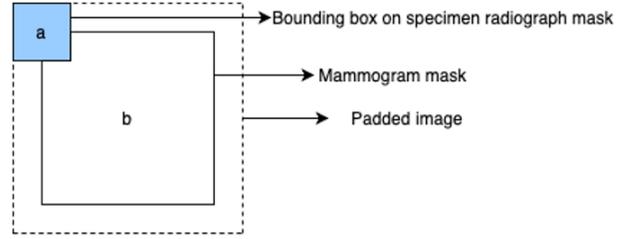

**Figure 3.** A schematic showing how images were padded prior to template matching.

### 2.5 Template matching

A template matching algorithm was applied to identify a patch in the magnification view with similar microcalcifications. The template matching algorithm [4] requires a template image ***a*** (here, the group of microcalcifications segmented from the specimen radiograph) and a target image ***b*** (i.e., a patch from the diagnostic mammogram) as inputs. The size of the template image was smaller than the size of the target image, since the template was the smallest rectangle containing a group of calcifications while the target image was the mammogram. Thus, for the computation to be performed on all (x, y) locations, the target image was padded with zero values in all directions by half the size of the template in each direction, as shown in **Figure 3**. The algorithm convolves the template a with image b by aligning the center of the template with each location (x, y) of the image and computing the inner product between the overlapping pixels.

$$f(x,y) = \sum_{x',y'} a(x',y') \cdot b(x + x', y + y') \qquad (Eq\ 1)$$

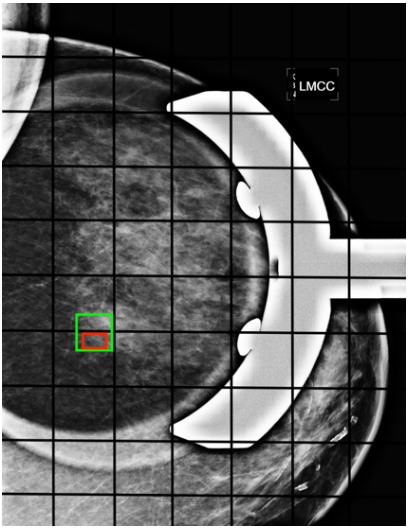

**Figure 4.** A schematic of the grid over the mammogram. The predicted patch is the one containing the red box, and the ground truth patch are the two patches with an overlap with the green box.

We utilized a cross-correlation metric (Eq 1), which is the inner product of each region and each template image, as a similarity score of that patch to the template. The resulting value is interpreted as a similarity score at each location. Similarity scores are calculated for each of the four templates. Since the dimensions of each template are different (due to varying sizes of the cluster of calcifications extracted from the specimen radiograph), the scores are not comparable across different cases. As such, we could not determine a universal threshold for choosing the best matching regions. Instead, the distribution of the scores generated for each location (x, y) of the image was plotted, and locations with a score higher than the 99th percentile of the score distribution were identified as the match.

### 2.6 Region scoring

The approach for scoring matched regions is illustrated in **Figure 4**. The top scoring patch (shown in red) that overlapped with the reference bounding box (shown in green) were counted as true positive. Patches that were not identified as matches and were not identified as biopsied regions by the human raters were considered true negatives (all other patches delineated by the grid). Metrics such as accuracy, precision, recall, specificity, and negative predictive value were calculated based on these definitions.

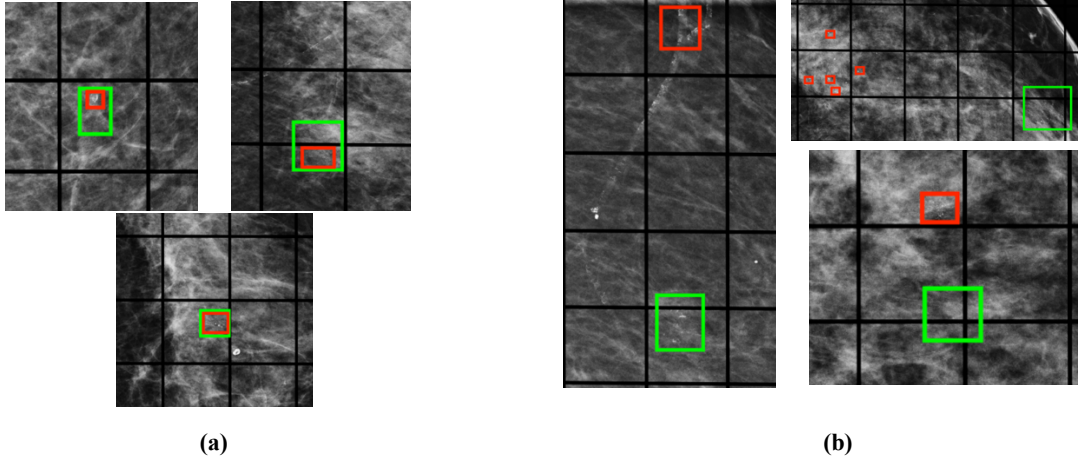

(a)                                                                                            (b)

**Figure 5.** Example results from our template matching-based approach. **(a)** True positive matches made by the algorithm and **(b)** failure cases.

## 3. RESULTS

We evaluated our approach using a test set of 80 cases, as described in Section 2.1. We also visually inspected the results as part of a failure analysis.

### 3.1 Performance on the test set

The magnified ML/LM-view images and the full CC and ML views were used to test our approach. **Table 1** summarizes the results. Our algorithm achieved consistent performance across different views. Given that the number of negative patches greatly outnumbered the number of positive patches, the accuracy and negative predictive value (NPV) was high, as expected. Precision and recall were reported at 0.66-0.69 and 0.58-0.63, respectively. The full ML/LM view achieved the highest precision and recall compared to other views.

| View | Accuracy | Precision | Recall | Specificity | NPV |
|---|---|---|---|---|---|
| Magnified ML | 0.99 | 0.66 | 0.61 | 0.99 | 0.98 |
| Full CC | 0.99 | 0.67 | 0.58 | 1.00 | 0.99 |
| Full ML/LM | 0.99 | 0.69 | 0.63 | 1.00 | 0.99 |

Table 1. Performance of our template matching-based approach on the test cases.

## 3.2 Visual inspection

We visually inspected the results across all test cases to understand where the algorithm succeeded and failed. **Figure 5** illustrates four examples of cases: two where the algorithm correctly identified the region and two where the algorithm failed. Our algorithm was likely to fail in several scenarios: 1) the appearance of the microcalcifications were the same in biopsied and non-biopsied regions; 2) the segmentation did not accurately capture the microcalcification shape; and 3) the biopsied regions extended across different patches.

## 4. DISCUSSION

We demonstrated an automated approach for calcification segmentation, clustering, and template matching approach that could be used to automatically identify the mammographic region where microcalcifications were taken during a core needle biopsy with a 0.66-0.69 precision and 0.58-0.63 recall. One of the most important challenges of our work, is the difference in the shape and the positions of the microcalcifications and the different nature of the specimen radiographs and the mammography images. Difference in the shape and position of the microcalcification and the surrounding tissue due to deformation of the breast during mammography, and changes in microcalcification shape due to tissue removal using a needle are the most apparent reasons. Moreover, our failure analysis identified the limitation of using a non-overlapping patch-based approach, which often divided regions of suspicious microcalcifications across multiple patches. Current methods rely on segmentation of the microcalcifications and using radiomics for classifying them, without spatially mapping the microcalcifications to the specimen radiograph. As such, assumptions are made as to how radiomic features calculated from the segmented microcalcifications relate to the pathology. Our approach is unique in that we attempt to spatialize localize the biopsied specimen to specific regions on the mammogram. Doing so, we hope to be able to more precisely relate quantitative image features extracted from mammograms to features extracted from digital pathology images. While our investigation is unique for mammograms, we note that Rusu et al. [5] examined a similar problem in magnetic resonance imaging (MRI) of the breast. They generated a 3D representation of the specimen by stacking multiple radiographs taken of each specimen and then spatially registering this volume to the breast MRI.

One significant limitation was the limited number of annotated diagnostic mammograms, magnification views, and specimen radiographs available for this analysis. Also, during the training of the microcalcification segmentation method, neither magnification views nor specimen radiographs were used during training. Refining the segmentation approach to work better on these types of views would mitigate the issue related to cases of microcalcifications where the shape is poorly captured. Finally, our initial approach to template matching assumes that a true positive match should occur in the image, given that the highest scoring template will always be considered a match. In practice, however, our algorithm may be used in cases where no microcalcifications exist in the image, resulting in the potential of false positives. While our algorithm could be applied only to biopsied cases, we intend to examine ways to determine negative results such as setting a minimum threshold on normalized template matching scores.

As part of future work, we will look at the images of patients with multiple biopsies from the same breast, attempting to match each specimen radiograph to its corresponding biopsy site. Furthermore, since the shape and the relative positions of the microcalcifications vary between the magnified view and the specimen radiograph, further investigation into the characteristics of each microcalcification detected will be performed to select the features that may be more unique to identify the corresponding region on the mammogram.

## REFERENCES

[1] Sun, Y.S., Zhao, Z., Yang, Z.N., Xu, F., Lu, H.J., Zhu, Z.Y., Shi, W., Jiang, J., Yao, P.P. and Zhu, H.P., 2017. Risk factors and preventions of breast cancer. *International journal of biological sciences*, *13*(11), p.1387.

[2] Marasinou C., Li B., Paige J., Omigbodun A., Nakhaei N., Hoyt A., and Hsu W., 2021. Segmentation of Breast Microcalcifications: A Multi-Scale Approach.

[3] Ester, M., Kriegel, H. P., Sander, J., & Xu, X. (1996, August). A density-based algorithm for discovering clusters in large spatial databases with noise. In KDD (Vol. 96, No. 34, pp. 226-231).

[4] Di Stefano, L., Mattoccia, S., & Mola, M. (2003, September). An efficient algorithm for exhaustive template matching based on normalized cross correlation. In 12th International Conference on Image Analysis and Processing, 2003. Proceedings. (pp. 322-327). IEEE.

[5] Rusu, M., Daniel, B. and West, R., 2019, March. Spatial integration of radiology and pathology images to characterize breast cancer aggressiveness on pre-surgical MRI. In Medical Imaging 2019: Image Processing (Vol. 10949, p. 109490Y). International Society for Optics and Photonics.